\documentclass{aa}
\usepackage{graphicx,times}
\begin{document}

   \title{First Light Measurements with the XMM-Newton Reflection Grating Spectrometers:
   Evidence for an Inverse First Ionisation Potential Effect and 
   Anomalous Ne Abundance in the Coronae of HR 1099}

   \author{A.C.~Brinkman
           \inst{1}
	   \and
	   E.~Behar
           \inst{2}
	   \and
	   M.~G\"udel
           \inst{3}
	   \and
	   M.~Audard
           \inst{3}
	   \and
	   A.J.F.~den Boggende
           \inst{1}
	   \and
	   G.~Branduardi-Raymont
           \inst{4}
	   \and
	   J.~Cottam
           \inst{2}
	   \and
	   C.~Erd
           \inst{5}
	   \and
	   J.W.~den Herder
           \inst{1}
	   \and
	   F.~Jansen
           \inst{5}
	   \and
	   J.S.~Kaastra
           \inst{1}
	   \and
	   S.M.~Kahn
           \inst{2}
	   \and
	   R.~Mewe
           \inst{1}
	   \and
	   F.B.S.~Paerels
           \inst{2}
	   \and
	   J.R.~Peterson
           \inst{2}
	   \and
	   A.P.~Rasmussen
           \inst{2}
	   \and
	   I.~Sakelliou
           \inst{4}
	   \and
	   C.~de Vries
           \inst{1}
          }

   \offprints{Bert Brinkman}

   \institute{Space Research Organization of the Netherlands, Sorbonnelaan 2, 3584 CA Utrecht,
The Netherlands
              \and
Columbia Astrophysics Laboratory, Columbia University, 550 West 120th Street, New York, 
NY 10027, USA
              \and
Paul Scherrer Institute, W\"urenlingen and Villigen, CH-5235 Switzerland
              \and
Mullard Space Science Laboratory, University College London, Dorking RH5 6NS, United Kingdom
              \and
Astrophysics Division, Space Science Department of ESA, ESTEC, 2200 AG Noordwijk, The Netherlands
             }

   \date{Received --; accepted --}

   \titlerunning{XMM-Newton RGS observations of HR\,1099}

   \abstract{
The RS\,CVn binary system HR\,1099 was extensively observed by the
{\it XMM-Newton} observatory in February 2000 as its first-light target.
A total of 570\,ks of exposure time was accumulated with the Reflection 
Grating Spectrometers (RGS). The integrated X-ray spectrum between 
5--35\AA\ is of unprecedented quality and shows numerous features attributed
to transitions of the elements C, N, O, Ne, Mg, Si, S, Fe, and Ni. 
We perform an in-depth study of the elemental
composition of the average corona of this system, and find that the
elemental abundances  strongly depend on the first ionisation potential 
(FIP) of the elements. But different from the solar coronal case, we find 
an {\it inverse FIP effect}, i.e., the abundances (relative to
oxygen) {\it increase with increasing FIP}. Possible scenarios, e.g., 
selective enrichment due to Ne-rich flare-like events, are discussed.
   }

   \maketitle

\section{Introduction}

The outer atmospheres of stars are of fundamental importance to the
chemical enrichment of interstellar space through stellar winds and the
ejection of energetic particles (Meyer 1985). Interestingly, the elemental composition
of the hot magnetic solar corona and the solar wind are markedly different
from the Sun's surface composition, which suggests that some fractionation
process in the denser layers of the solar atmosphere must selectively
enrich the corona with particular elements. 
Various observations of the solar corona, the solar wind, solar energetic
particles, and cosmic rays have established that elements with a first
ionisation potential (FIP) below 10\,eV are overabundant by factors of up
to 4 with respect to the solar photosphere, unlike elements with a high
FIP that maintain their photospheric abundance (Feldman 1992, see also 
Meyer 1985). 
This is usually referred to as the ``FIP effect". 
However, the physics of this
fractionation mechanism is poorly understood. Comparative abundance
studies of stars other than the Sun may shed light 
on the underlying physics.

\begin{figure*}[ht]
\resizebox{12cm}{!}{\includegraphics[angle=0, clip, bb=43 41 716 525]{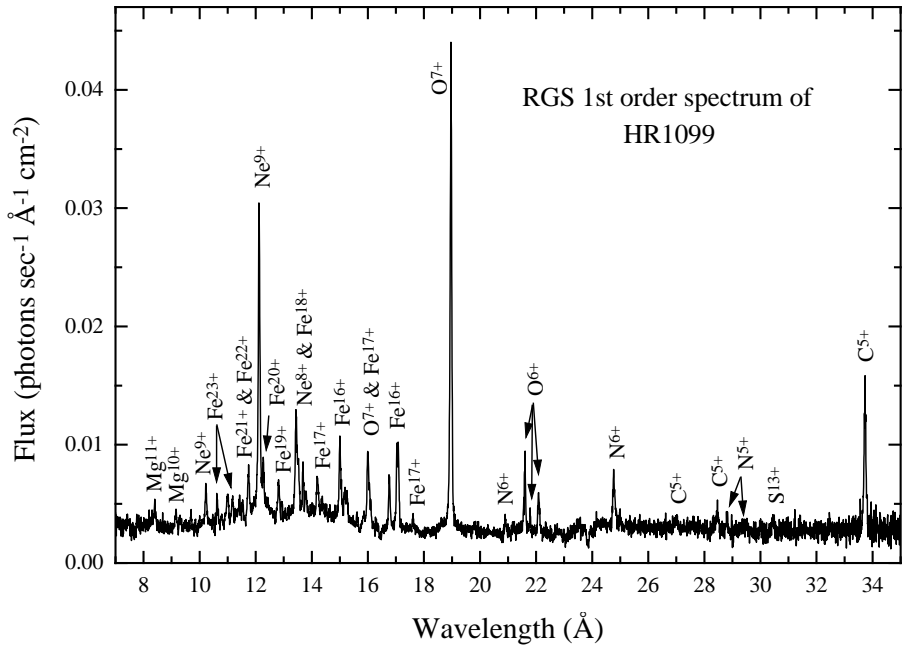}}
\hfill
\parbox[b]{55mm}{
\caption{First order spectrum of HR\,1099 as measured by the RGS
instruments on board XMM-Newton. The most prominent spectral features are
indicated just above the spectrum by the corresponding emitting ion.}
\vspace{1cm}
}
\end{figure*}

The X-ray regime contains
a rich forest of atomic spectral lines emitted by many chemical species at
ionisation stages corresponding to plasma temperatures between $\sim$1 and 30
Million K.  So far, however, attempts to determine the composition of
stellar coronae other than the Sun's from X-ray observations 
(Antunes et al.\ 1994, White 1996, Drake 1996)
have produced results that are considered to be ambiguous, because the low
spectral resolution available from even the best nondispersive detectors
has not permitted the separation of atomic emission lines arising from
different elements. Often, subphotospheric abundances were claimed.
Studies of stellar coronal abundances with the EUVE satellite 
have revealed solar-like FIP effects in some, but not all of the stars
studied (Drake et al.\ 1995, 1997). 
The high-resolution grating spectrometers on board the
recently launched American Chandra and European XMM-Newton satellites now
provide a new opportunity to unambiguously determine both the elemental
abundances and the thermal structure of stellar coronae through explicit
measurement of emission line and continuum intensities in the X-ray band.
We are using grating spectroscopy on XMM-Newton to perform such analyses
for the first time. The target chosen for the present observation is HR\,1099, 
one of the X-ray brightest and most magnetically active stellar
systems, which was observed early in the programme. 
HR\,1099 has been the target of previous EUVE and low-resolution X-ray spectroscopic
investigations (e.g.\ Pasquini et al.\ 1989, Griffith \&\ Jordan 1998).
Ayres et al.\ (2001) qualitatively compare the Chandra HETG spectra of
HR\,1099 and Capella, the latter being a somewhat cooler and less active coronal
source. A companion paper (Audard et al.\ 2001) to the present work discusses 
spectral variability in HR\,1099, based on selected XMM-Newton RGS data sets.

\section{Observations and Analysis}

The XMM-Newton X-ray observatory incorporates a payload with
two identical high-resolution Reflection Grating Spectrometers (RGS;
den Herder et al.\ 2001) with
a spectral resolution of about 0.06\,\AA\ FWHM. The spectral band covered by
the RGS ranges from 5 to 35\,\AA. This range contains strong lines of C, N,
O, Ne, Mg, Si, S, Ca, Fe, and Ni. The clearly detected eight consecutive
charge states of Fe are of crucial importance for the present analysis. 
HR\,1099 was observed in January and February 2000
for a total of 570\,ksec yielding approximately 
1.3 million source counts. The spectrum was extracted by counting events
inside a narrow spatial window along the dispersed CCD image, and then by
using the CCD energy resolution to separate the spectral orders. The
extracted total first-order spectrum between 7 and 35\,\AA\ is shown in 
Fig.~1. 

\section{Results}

In optically thin coronal plasmas, the power of a particular spectral line depends on the
collisional and radiative transition rates. In this work,
these quantities are calculated by means of the relativistic Hebrew University Lawrence
Livermore Atomic Code (Bar-Shalom et al.\ 1998). Owing to the
high spectral resolution, we can directly scale the contribution of each ion species to match
the observation. The atomic-level models are each calculated for a single electron
temperature, T$_{\rm max}$, at which the relevant ion has its maximum abundance. 
Fig.~2 shows the theoretical spectrum from 7--18\,\AA\ (red curve), which
also includes a phenomenological continuum component (blue)  
corresponding to a bremsstrahlung spectrum (which dominates over
any other continuum components at the temperatures in question), compared with
the HR\,1099 data points (black). As can be seen, the
agreement is very good. The majority of lines in this region are due to
highly ionised Fe as demonstrated by the individual ion curves at the
bottom of the figure. The contribution of each ion species yields 
a direct measurement of the amount of coronal gas
in the conditions appropriate for that ion.  

With the distance d (= 29\,pc)
to HR\,1099, the calculated line power and the measured line flux can be
used to obtain the emission measure EM (in cm$^{-3}$), which is a product of
the electron density and the H-ion density, integrated over the emitting
volume. We
assume that each ion emits mostly around the temperature T$_{\rm max}$. 
This leads to a first approximation of the emission measure distribution.
This distribution is sufficiently accurate for our purposes if no
steep gradients are present.
Finally, in order to plot the EM, we use the ionic abundances from 
ionisation balance calculations (Arnaud \&\ Raymond 1992, Mazotta 1998) and assume,
as a first step, elemental abundances of the solar photosphere.
The resulting EM as a function of the electron temperature T$_{\rm e}$ is shown in 
Fig.~3 for ions of C, N, O, Ne, Mg, S, Fe, and Ni. These ions sample the 
HR\,1099 hot coronae in the T$_{\rm e}$ range from kT$_{\rm e}=100$\,eV to 1500\,eV and
illustrate the monotonic increase of the EM in this range. The errors on
the temperature for each ion represent the range of T$_{\rm e}$ for which the ion
fraction exceeds 50\%\ of its maximum value, except for the He-like ions
O$^{6+}$, Ne$^{8+}$, and Mg$^{10+}$, for which line ratio methods 
(Gabriel \&\ Jordan 1969) are used to
determine T$_{\rm e}$ more precisely and the bars reflect a 25\%\ uncertainty.

\begin{figure*}[ht]
\resizebox{12cm}{!}{
\includegraphics[angle=270, origin=c, clip, bb=80 54 568 752]{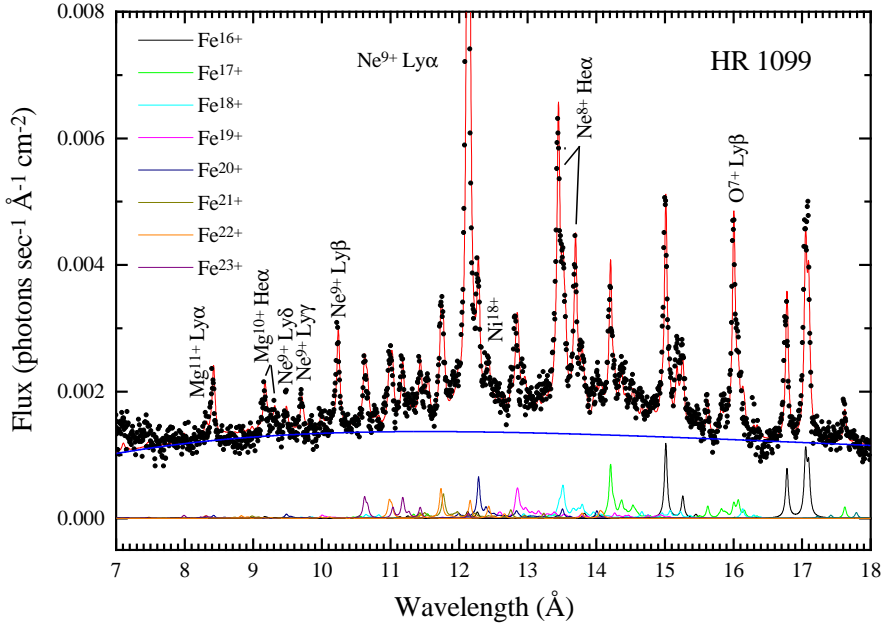}}
\hfill
\parbox[b]{55mm}{
\caption{Observed HR\,1099 flux spectrum (black dots) compared with the
total calculated spectrum (red curve). The inferred thermal bremsstrahlung
continuum is represented by the blue curve. The eight separate plots at
the bottom show the relative individual-ion contributions of Fe$^{16+}$ to Fe$^{23+}$
(not to scale). The
non-Fe lines (O, Ne, Mg, and Ni) are indicated explicitly.}
\vspace{2.7cm}
}
\vspace{-1.5cm}
\end{figure*}

\begin{table*}[t]
\caption{Abundance estimates$^a$ of the HR\,1099 coronae, Solar photosphere$^b$ and Solar corona$^c$ relative to O}
\begin{tabular}{ccccccc}
\hline
Element & FIP   & HR\,1099  & Solar  & Solar  & Solar         & HR\,1099      \\
        & (eV)  & coronae   & photo- & corona & corona/Solar  & coronae/Solar \\
        &       & (present) & sphere &        & photosphere   & photosphere   \\
\hline
Ne      & 21.56       & 0.57      & 0.15   & 0.15   & 1.0  & 3.8 \\
N       & 14.53       & 0.17      & 0.12   & 0.13   & 1.08 & 1.4 \\
O       & 13.61       & 1.0       & 1.0    & 1.0    & 1.0  & 1.0 \\
C       & 11.26       & 0.28      & 0.47   & 0.50   & 1.06 & 0.59 \\
S       & 10.36       & 0.0085    & 0.019  & 0.024  & 1.26 & 0.45 \\
Fe      & 7.87        & 0.0094    & 0.038  & 0.16   & 4.21 & 0.25 \\
Mg      & 7.64        & 0.043     & 0.045  & 0.18   & 4.0  & 0.95 \\
Ni      & 7.63        & 0.00070   & 0.0021 & 0.0089 & 4.24 & 0.33 \\
\hline
\multicolumn{6}{l}{\footnotesize $^a$\,Error estimates are 20\%\ for all numbers.} \\
\multicolumn{6}{l}{\footnotesize $^b$\,Feldman (1992).} \\
\multicolumn{6}{l}{\footnotesize $^c$\,Feldman et al.\ (1992).} \\
\end{tabular}
\end{table*}

Looking only at the eight Fe data points in Fig.~3, the EM distribution
appears to be approximately a smooth power law over the relevant
temperature range as indicated by the solid line. The use of Fe alone
provides a representation of the EM behaviour independent of elemental
abundance. Note that the power-law EM distribution obtained for Fe is
strikingly parallel to the EM trends suggested by the data points for O,
Ne, and Mg. However, the distribution is vertically offset for these other
elements. The corrections needed in order to settle these offsets allow us
to estimate the abundances relative to one another. The high degree of
overlap in the T$_{\rm e}$ ranges of formation, especially for the Ne, Mg, Fe, and
Ni ions, makes the sense of the relative abundance determinations
unambiguous. 
Table~1 presents our derived approximate abundances with the
solar photospheric and solar coronal abundances. The latter
should be considered as average abundances in various solar
coronal features in which the individual abundances may vary 
somewhat (H\'enoux 1995). Similarly, our HR\,1099 results are average
coronal abundances during our observations. We note that we 
give abundances relative to the well-determined, abundant, 
and well-behaved (on the Sun) high-FIP element O. Relative 
abundances are appropriate for the study of possible dependencies 
on the FIP. They also avoid complications with the detailed modeling 
of the continuum. We plot in Fig.~4 the coronal enrichment (relative 
to solar photospheric abundances) for the HR\,1099 coronae as a 
function of the FIP.

We note in passing that the precise composition of the HR\,1099 
photosphere (except for a recent measurement of the Fe abundance 
of both HR\,1099 components, Randich et al.\ 1994) is unknown.
It is expected that overall, the photospheres of nearby field stars 
such as HR\,1099 have a mixture of elements similar to the Sun.

\begin{figure*}[t]
\resizebox{12cm}{!}{\includegraphics[angle=-90, origin=c, clip, bb=85 51 580 715]{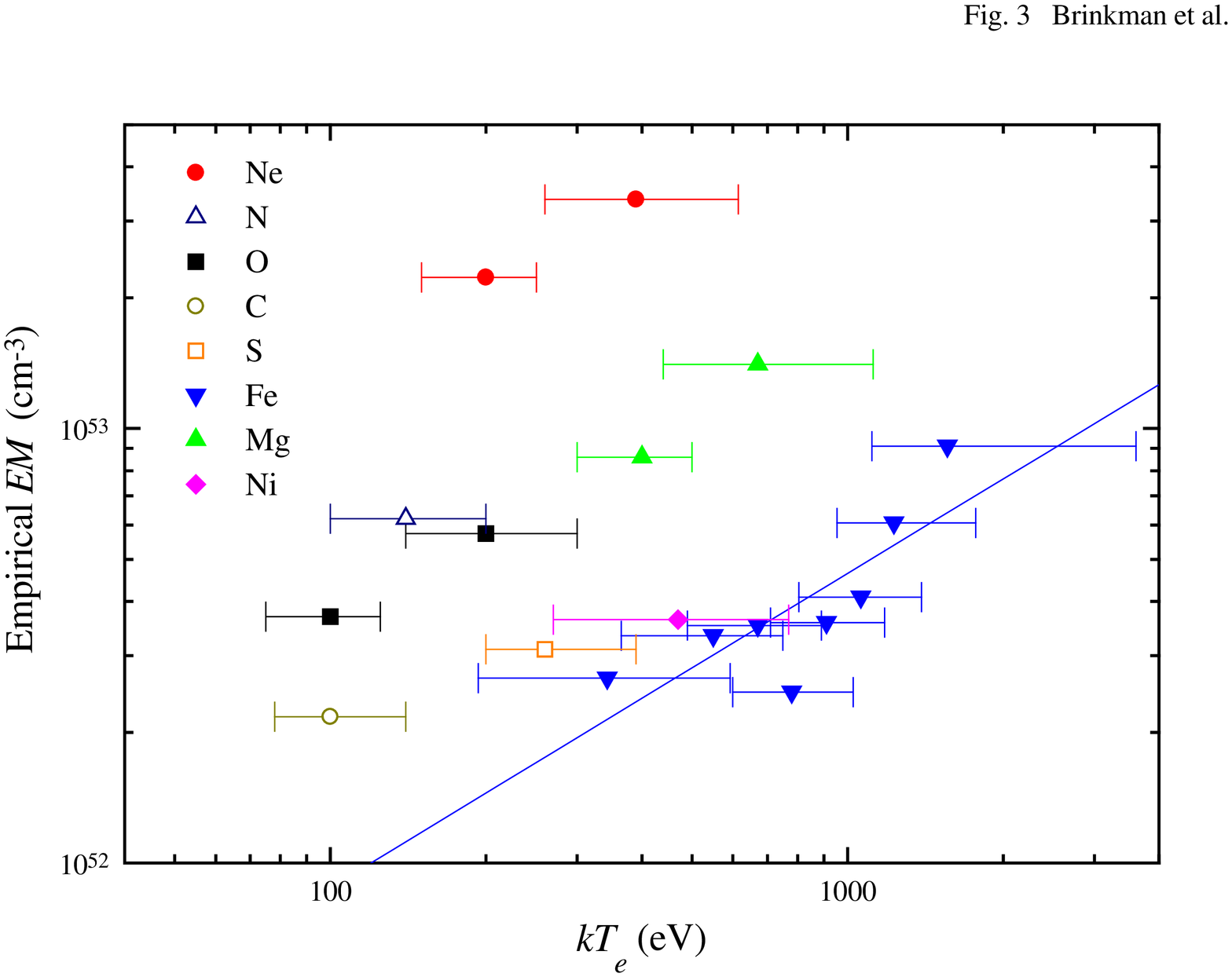}}
\hfill
\parbox[b]{55mm}{
\caption{The empirical emission measure for: O$^{6+}$, O$^{7+}$, Ne$^{8+}$, Ne$^{9+}$, 
Mg$^{10+}$, Mg$^{11+}$, Fe$^{16+}$ through Fe$^{23+}$, Ni$^{18+}$, C$^{5+}$, N$^{6+}$ 
and S$^{13+}$ (in order from left to right for each element). The straight line is the best-fit 
power-law for the Fe ions alone.}
\vspace{2.7cm}
}
\vspace{-1.5cm}
\end{figure*}

\begin{figure*}[t]
\resizebox{12cm}{!}{\includegraphics[angle=-90, origin=c, clip, bb=85 51 580 715]{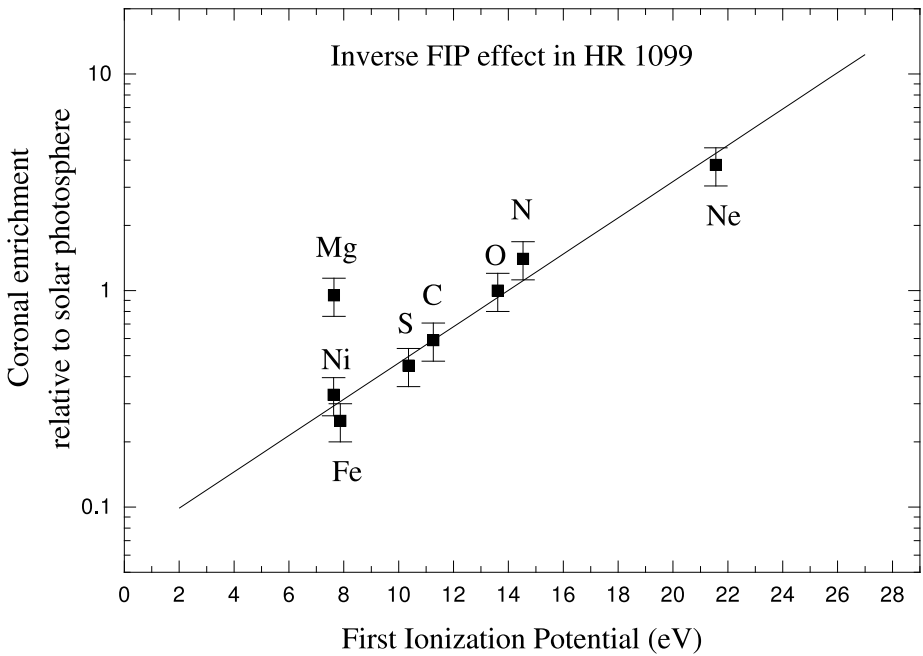}}
\hfill
\parbox[b]{55mm}{
\caption{Enrichment of elemental abundances in the HR\,1099 coronae relative to (solar) 
photospheric values, plotted as a function of the first ionisation potential (FIP).
All values are given error bars of 20\%. Note the increase with FIP, exactly opposite 
the trend observed in the average solar corona.}
\vspace{2.7cm}
}
\vspace{-1.5cm}
\end{figure*}

\section{Discussion}

The Ne/O ratio found here for the HR\,1099 coronae is unusually high
compared with the solar photospheric value, while the N/O ratio is
also enhanced and the Mg/O ratio is close to solar photospheric.
The abundance ratios for the remaining elements are well below their solar
photospheric values. Compared with the respective ratios of the solar
corona, the contrast is even sharper. In fact, HR\,1099 reveals
systematically a reversed FIP effect, which is illustrated in Fig.~4; 
the coronal enrichment increases
monotonically with the FIP. The only exception may be Mg (low FIP). 
Ne is strongly enhanced, by a factor of 3.8 compared to the solar photosphere.
This enhancement is reminiscent of a
similar anomaly observed in a subset of solar flares (Schmelz 1993, Murphy et al.\ 1991). 
Interestingly,
flares are thought to be of great importance to the heating of apparently
quiet coronae of active stars (Doyle \&\ Butler 1985, G\"udel 1994, Audard et al.\ 2000) 
and have been suggested to be pivotal for the elemental fractionation process as well 
(Wang 1996, G\"udel et al.\ 1999, Mewe et al.\ 1997). Continuous mixing
of photospheric and coronal material in low-lying loops through frequent
flares could suppress the solar-like FIP effect (Feldman \&\ Widing 1990) in these stars
altogether.

It is generally believed that the fractionation is the result
of diffusive processes across magnetic field lines somewhere in the
chromosphere, where the ionisation fraction differs substantially among
elements of high or low FIP (von Steiger \&\ Geiss 1989). 
Low FIP elements can then be accelerated
into the quiescent corona, leaving an excess of high FIP elements in the
chromosphere (Schmelz 1993). If this chromospheric layer can be accessed by flares
(Schmelz 1993), then a high-FIP-rich plasma may be heated and lifted into dense coronal
X-ray loops giving rise to the reversed coronal FIP effect observed in 
HR\,1099. In any case, it is clear that the simple mechanism involving
steady-state diffusion, which has been proposed for the quiet solar
corona, does not operate in more active stars such as HR\,1099. Since cosmic
rays do show a low FIP enhancement, it is unlikely that such active stars
can be major contributors to the cosmic ray flux. The wide survey of
stellar coronae planned with the RGS on XMM-Newton will provide more
general answers to the questions raised by these observations.

\begin{acknowledgements}
We thank the many teams in industry and at the scientific institutions and
ESA, for building such an excellent spacecraft and payload - and operating
it. We are particularly grateful to Mr. Robert Lain\'e and his project team
at ESA for leading the project. We thank the XMM-Newton SOC staff for
making the commissioning phase data available to us.  SRON is financially
supported by the Netherlands Organization for Scientific Research (NWO).
The Columbia University team acknowledges generous support from the
National Aeronautics and Space Administration. The PSI group is supported
by the Swiss Academy of Natural Sciences and the Swiss National Science
Foundation (grants 2100-049343 and 2000-058827). 
MSSL acknowledges support from the Particle Physics and
Astronomy Research Council. This work is based on observations obtained with
XMM-Newton, an ESA science mission with instruments and contributions directly
funded by ESA member states and the USA (NASA).
\end{acknowledgements}

\end{document}